\documentclass[aps,prl,showpacs,floats,10pt,twocolumn]{revtex4}
\usepackage{graphicx}
\usepackage{amsfonts}
\begin{document}
\def\E{{\bf E}}

\def\T {|q\rangle \langle q|} \def\dip{{3 |{\bf r}\rangle \langle {\bf r}| -1
    \over 4 \pi r^3}}

\def\P{{\bf P}}

\def\D{{\bf D}}

\def\rr{{\bf r}}

\def\dV{{\rm d^3}{\bf r}}

\def\qq{{\bf q}}

\def \div{ {\rm div}\, } \def \grad{ {\rm grad}\, } \def \curl{ {\rm curl}\, }

\def\p{{\bf p}}

\def\rhat{\hat {\bf r}}

\def \dv{\; d^3\rr}

\title{Simulating nanoscale dielectric response} \author{A. C. Maggs}
\affiliation{Laboratoire de Physico-Chime Th\'eorique, UMR CNRS-ESPCI 7083, 10
  rue Vauquelin, 75231 Paris Cedex 05, France.}  \author{R. Everaers}
\affiliation{Max-Planck-Institut f\"ur Physik komplexer Systeme, N\"othnitzer
  Str.  38, Dresden, Germany }
\begin{abstract}
  We introduce a constrained energy functional to describe dielectric
  response. We demonstrate that the local functional is a
  generalization of the long ranged Marcus energy.  Our re-formulation
  is used to implement a cluster Monte Carlo algorithm for the
  simulation of dielectric media.  The algorithm avoids solving the
  Poisson equation and remains efficient in the presence of spatial
  heterogeneity, nonlinearity and scale dependent dielectric
  properties.
\end{abstract}
 \maketitle

The response of dielectric materials leads to important modifications of the
bare electrostatic energy of charges \cite{born}. Even on {\em microscopic}\/
scales many important qualitative insights result from the application of the
{\em macroscopic}\/ laws of electrostatics.  For example, a large part of the
solvation free energy of an ion in water is understood in terms of the Born
self energy of an ion, $q^2/8 \pi a \epsilon$ where $\epsilon$ is the
macroscopic dielectric constant of the solvent and $a$ an atomic scale;
similarly many essential features of ion channels can be explained by a
continuum description of high dielectric channels through low dielectric
membranes~\cite{parsegian,sklovskii}. 


On the nm-scale the dielectric response is in general non-local \cite{dolgov},
\begin{math}
  \D_\rr = \int \epsilon_{\rr,\rr'} \E_{\rr'}\dv',
\end{math} 
where $\D$ is the electric displacement and $\E$ the electric field.
In Fourier space the field energy of a
homogeneous fluid is given by
\begin{equation}
U_{elec}
=\sum_q \epsilon(q) {\E^2(q) \over 2} \label{UD}.
\end{equation}
where $\epsilon(q)$ is now a scale dependent dielectric constant with
\begin{math}
  \D({q}) = \epsilon(q) \E({q})
\end{math}.
If we take water as an example \cite{bopp} we learn that starting from
$\epsilon\sim 80$ at $q=0$, the dielectric constant of water drops to
$\epsilon=20$ at a wavelength which is comparable to the Bjerrum length of a
monovalent ion (7 \AA); below a wavelength of $\sim$5\AA\, $\epsilon$ diverges
and becomes {\em negative}. Moreover, many situations of practical interest
require the inclusion of non-linear effects such as dielectric saturation
or surface ordering and alignment \cite{ball_nat_03}.

Recently, we introduced an algorithm which permits the calculation of
electrostatic interactions  in heterogeneous {\sl local}\/
dielectric media without solving the Poisson equation \cite{acm1}.
The idea is to generate long ranged electrostatic interactions {\sl
  dynamically}\/ via local interactions between charges, the medium
and the electric field. On first sight, it seems straightforward to
replace the field energy
\begin{math}
  U =  \int \D^2_{\rr}/2\epsilon_\rr \dv
\end{math}
used in \cite{acm1} by eq.~(\ref{UD}). However, this approach is fundamentally
flawed: the method would allow one to treat {\em unphysical}\/ systems with
$0<\epsilon(q)<1$ \cite{dolgov} while becoming unstable for real materials in
which the dielectric constant becomes negative.

In this letter we present a generalization of the Marcus energy for
polarizable media~\cite{marcus} written in terms of the true, independent
thermodynamic variables in the problem \cite{dolgov,kirzhnits}: the electric
polarization of the medium, $\P$, and the displacement, $\D$.  We show that
our formalism is capable of reproducing the full range of the linear
dielectric response seen in nature including negative dielectric constants.
\cite{dolgov,kirzhnits}.  In addition, we demonstrate that the same
techniques can be used to study non-linear or polar dielectrics at
essentially identical computational cost.  The key feature is the use of a
cluster algorithm~\cite{statphys} for the equilibration of the field degrees
of freedom.  Similarly to related problems for classical \cite{swendson} and
quantum spins \cite{alet} the numerical effort per autocorrelation time can be
reduced from $O(L^z)$ sweeps with $z > 0$ to $O(L^0)$ sweeps 
The prefactor in this scaling depends on the dielectric properties.

We now introduce the energy functional. There are two contributions to
the energy of a dielectric medium.  Firstly, the energy density
$\E^2/2 = (\D-\P)^2/2$ of the electric field; secondly, an elastic
polarization energy, $G(\P)$, due to short ranged interactions between
molecules.  We start by expressing $G$ as a general quadratic function
of $\P$ with a {\sl short ranged kernel}\/ $K_{\rr,\rr'}$ which we
describe more fully below.  In a heterogeneous system $K$ varies from
place to place in the sample; it contains {\sl all}\/ the material
properties. Thus,
\begin{equation}
  U= \int {(\D_\rr-\P_\rr)^2\over 2} \dv 
  + \int {\P_\rr K_{\rr,\rr'} \P_{\rr'} \over 2} \label{UP} \dv \dv'
\end{equation}
We use units with $\epsilon_0=1$ and periodic boundary conditions.
Furthermore, $\D$ is constrained by Gauss' law, $\div \D -\rho=0$.  In
the following we demonstrate the equivalence of our formalism to the
standard theory of dielectric media \cite{marchi,felderhof,fulton}.

We first work at zero temperature; this will allow us to calculate the
relationship between the dielectric constant $\epsilon(q)$ defined in
eq.~(\ref{UD}) and the kernel $K(q)$ in eq.~(\ref{UP}).  We minimize
eq.~(\ref{UP}) subject to the constraint of Gauss' law with the help of a
Lagrange multiplier $\phi$.  We consider the stationary points of the
functional
\begin{equation}
A= U - \int \dv \;\phi(\div \D -\rho) \label{A}
\end{equation}
We will pass rather freely between the full integral form, eq.~(\ref{UP}) and
an operator notation in which all components of a field are grouped in a
single vector and $K$ is a matrix. The variational equations are:
\begin{eqnarray}
\delta \phi&:&\quad \div \D -\rho =0 \nonumber \\
\delta \P&:&\quad \P-\D+ K\P=0 \label{DP}\\
\delta \D&:&\quad \D-\P + \grad \phi=0 \nonumber
\end{eqnarray}
We indeed find from the variational equations that if we define $\E \equiv \D-\P$
then the relation between the polarization and the electric field is
\begin{math}
  \P= \chi \E
\end{math}, where  $\chi=K^{-1}$ is the susceptibility to the electric
field, $\E=-\grad \phi$.  We now solve eq.~(\ref{DP}) for $\P$ in terms of
$\D$ and substitute in eq.~(\ref{UP}). We find eq.~(\ref{UD}) with
\begin{equation}
  \epsilon(q)= 1+K^{-1}(q) \label{eps}
\end{equation}
We have reproduced all the standard relations between $\E$, $\P$,
$\phi$ and $\D$ of electrostatics, as well as the energy
eq.~(\ref{UD}) which should be the {\em minimum}\/ of the functional
eq.~(\ref{UP}).  Similarly, we can show that our Ansatz is equivalent
to the Marcus polarization functional \cite{marcus,felderhof,marchi}:
we use eq.~(\ref{DP}) to eliminate the constrained field $\D$ and
express eq.~(\ref{UP}) in terms of the polarization $\P$ and the {\em
  bare}\/ electric field
\begin{math}
  \E_0 = -\grad \phi_0
\end{math}
generated by the free charges in vacuum:
\begin{math}
  \nabla^2 \phi_0 = - \rho
\end{math}:

\begin{eqnarray}
U_p &=&  {1\over 2} \int \dv \dv'\;   {\div \P_\rr \div \P_{\rr'} \over |\rr-\rr'|} 
- \int \E_0 \cdot \P \dv \nonumber \\
&+& {1\over 2} \int   \P_\rr K_{\rr,\rr'} \P_{\rr'} \dv\dv' 
+\int { \E_0^2 \over 2} \dv
\label{marcus}
\end{eqnarray}
It is instructive to rewrite Eq.~(\ref{marcus}) in the operator notation:
\begin{eqnarray}
U_p &=& {1\over 2} \P (T+K) \P   - \E_0 \cdot \P 
+ {\E_0^2 \over 2} \label{dipoleenergy}
\end{eqnarray}
$T$ denotes 
\begin{math}
   \T  \rightarrow -{3 |{\bf r}\rangle \langle {\bf r}| 
- 1 \over 4 \pi    r^3} + {I\over 3} \delta({\bf r}) \label{project}
\end{math} in real space.  The field $\P$ interacts with itself via
the long ranged dipole-dipole potential.

We now consider the stability of the functional eq.~(\ref{UP}) for $q \ne 0$,
in order to check that the energy is a true minimum, not just a stationary
point.  The longitudinal constraint of eq.~(\ref{A}) freezes $\D$; only
fluctuations of $\P$ are free.  The coefficient of ${\P} ^{2} $ in
eq.~(\ref{UP}) is $V_q(\P) = \P (1+K)\P /2$. In order for fluctuations in $\P$
to be bounded (so that the ground state is stable) we require that eigenvalues
of $(1+K)$ are positive.  Now express $V_q(\P)$ in terms of the field $\E$ and
eliminate $K$ for $\epsilon$ using eq.~(\ref{eps}):
\begin{equation}
V_q(\E) = {\epsilon(q)(\epsilon(q)-1)\over 2} \E^2\quad >0 \label{VE} 
\end{equation}
 Stability implies that $\epsilon(q)(\epsilon(q)-1)>0$, so that $\epsilon \geq 1$ or
$\epsilon \leq 0$.  The expression eq.~(\ref{VE}) is indeed that given in
\cite{kirzhnits} for the effective potential 
of the electric field.  Stable modes with $\epsilon(q) <0$ (such as those seen
in water) correspond to $-1<K(q)<0$.  We conclude that our constrained energy
eq.~(\ref{UP}) is capable of reproducing the full range of dielectric response
seen in nature and leads to the correct band of forbidden response functions
$0<\epsilon(q)<1$ where the system becomes thermodynamically unstable: in
electrostatics there can be no equivalent of diamagnetic response, which might
seem plausible from the conventional energy eq~(\ref{UD}).

We now generalize to finite temperatures and study the partition function
\begin{equation}
Z= \int d\P\; d\D\; e^{-\beta U} \prod_{{\bf r}} \delta(\div \D(\rr) 
-\rho(\rr)) \label{Z}
\end{equation}
We shall integrate over either $\P$ or $\D$ to find the finite temperature
generalizations of eq.~(\ref{UD}) and eq.~(\ref{marcus}): The $\P$ degrees of
freedom are Gaussian, when we integrate over them we find the constrained
partition function
\begin{equation}
Z= {1\over  \sqrt{|1+K|}}\int d\D 
e^{-\beta \int {\D^2\over 2 \epsilon} \dv} \prod_\rr \delta( \div \D -\rho) \label{ZD}
\end{equation}
where we have dropped unimportant numerical prefactors.  This constrained
integral over $\D$ was studied in detail in \cite{acm1}.  When dielectric
properties are local, so that $K(\rr,\rr')= \kappa(\rr) I \delta(\rr-\rr')$,
eq.~(\ref{ZD}) describes particles interacting through an electrostatic
potential which is a solution to the Poisson equation,
\begin{math}
  \div (\epsilon \grad \phi) =-\rho \label{poisson}
\end{math}, with $\epsilon(\rr)= 1+1/\kappa(\rr)$.
  In addition it gives the Keesom potential between
fluctuating, classical dipoles.

Rather than integrating over $\P$ we integrate over $\D$ in eq.~(\ref{Z}): the
$\delta$-function constraint is imposed using the identity
\begin{math}
  2 \pi \delta(x) = \int e^{i \phi x}\, d \phi
\end{math}.
The integral over $\D$ is then Gaussian, as is that subsequently performed
over $\phi$.  We find
\begin{math}
  Z= \int d\P e^{-\beta U_p}
\end{math}.  $U_p$ is the energy of eq.~(\ref{marcus}).  This is our
principal formal result. It shows that with the energy functional
eq.~(\ref{UP}) integrating over the constrained field $\D$ generates results
which are equivalent to using the long ranged Marcus functional
eq.~(\ref{marcus}).

The treatment of the stability of the field $\P$ requires generalization at
finite temperature: We must distinguish between the longitudinal and
transverse parts of the operator $K$ when $q \ne 0$: $K_{l|t}$. 
From eq.~(\ref{UP}) we calculate the fluctuations of the polarization
field.  Longitudinal fluctuations give
\begin{math}
  \beta S({\bf q})=\beta \langle \delta \P\cdot \delta \P\rangle _l({\bf q}) =
  {1/ (1+K_l({\bf q}))}
\end{math}
whereas for the two transverse modes
\begin{math}
  \beta \langle \delta \P\cdot \delta \P \rangle _t ({\bf q})= {2/ K_t({\bf
      q})}
\end{math}, so that the eigenvalues of $K_t$ must also be positive.
Combining these two expressions we find the fluctuations in $\P$ for small
$q$:
\begin{eqnarray}
\beta \langle \delta \P\cdot \delta \P \rangle_{q\rightarrow 0} 
&=& {1\over (1+K_l)} + {2\over K_t} \rightarrow  {(2 \epsilon+1)(\epsilon-1)\over \epsilon} \nonumber \\
\beta \langle \delta \P^2 \rangle_{q=0} &=& {3 \over K(0)} =3(\epsilon-1)\label{k0}
\end{eqnarray}
where we have used the fact that $K$ becomes isotropic at small $q$.  These
expressions are familiar from the Kirkwood theory of dielectrics
\cite{kirkwood} 
and will be used to extract dielectric constants from our simulations.
Using these expressions is always numerically challenging since they
link the dielectric properties to {\sl fluctuations}\/ in the
polarization field.  Accurate results require simulations which are
many times longer than the equilibration time.

\begin{figure}[htb]
  \includegraphics[scale=.55] {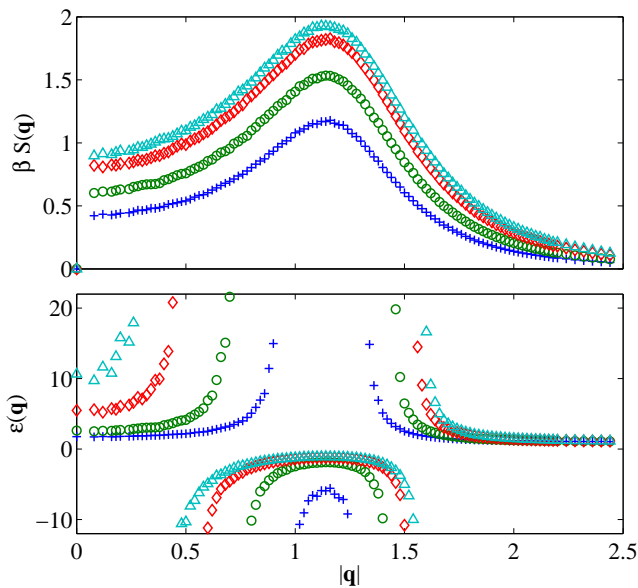}
\caption{Dielectric constant deduced from polarization fluctuations for the
  dielectric eq.~(\ref{nonlinear}).  $L=64$. Data plotted as a function of
  $|{\bf q} |$ with $| {\bf q}|^2 = 2 \sum_i (1-\cos q_i)$.
  $(\gamma,\kappa_l)$ are $+$ (5,-2.3), $\circ$ (2.5,-1.5), $\Diamond$ (1,
  -1.0), $\triangle$ (0.5,-0.9).  Four days simulation per curve on a
  Pentium-4 processor.
\label{epsq}
}\end{figure}

By choosing appropriate kernels $K$ we can introduce semi-microscopic
models of dielectric media with a great variety of dielectric
properties. A systematic approach is to use a Landau-Ginsburg
expansion of the free energy for the polarization
field~\cite{kornyshev}.  The lowest order terms (corresponding to a
linear theory $G=\P K\P/2$) are given by

\begin{equation}
G={1\over 2}\int \left \{ 
\kappa  \P^2 +  \kappa_t (\curl \P)^2 + \kappa_l (\div \P)^2 \right \} \dv
\label{linear}
\end{equation}
Eigenvalues of $K_l$ are $(\kappa+ \kappa_l q^2)$, transverse eigenvalues are
$(\kappa+\kappa_t q^2)$. The dispersion laws split at $O(q^2)$. Stability
implies that $\kappa$, $\kappa_l$ and $\kappa_t$ must all be positive. From
eq.~(\ref{eps}) the dielectric constant is $\epsilon(q)=\{1 +1/(\kappa +
\kappa_l q^2)\}$. It falls off monotonically with wavevector.

Until now we have worked with linear media, which as we have shown
analytically reproduce the standard theory of dielectric media.
However our approach is {\sl not}\/ limited to linear models.  Consider
soft Langevin dipoles where the length constraint is imposed by an
energy function
\begin{math}
  G= \int \gamma (\P^2-P_0^2)^2 \dv.
\end{math}
Dipoles with long ranged interactions described by the operator $T$ in
eq.~(\ref{dipoleenergy}) have long been used \cite{warshel2} to
describe polar solvents. In contrast to linear models, they exhibit a
saturation of the polarization degrees of freedom when interacting
with strongly charged objects.  To generate, in addition, scale
dependent dielectric effects we include derivative terms in the free
energy:
\begin{equation}
G=  \int \left\{ \gamma (\P^2-P_0^2)^2 + {\kappa_{l} (\div \P)^2\over 2} + {\alpha 
(\grad \div \P)^2 \over 2} \right\}
  \dv
\label{nonlinear}
\end{equation}
We are particularly interested in the case $\kappa_{l} <0$ in order to produce
systems with $\epsilon(q)<0$, together with $\alpha>0$, necessary for
stability at large wavevectors.  At interfaces other contributions to $G$ such
as $ \int \grad \kappa \cdot \P \dv$ select a preferred direction for the
polarization and can be used to study ordering by surfaces. We leave such
considerations, however, to future work. In our first simulations we used
$\alpha=-0.4 \kappa_{l}$, $P_0^2= 7.5 /\gamma$.

We discretize the fields on a simple cubic lattice of side $L$ so that
$\E$ and $\P$ are associated with the $3L^3$ links.  The constrained
field $\D$ is sampled by a cluster (worm) algorithm
\cite{alet,statphys} invented to study quantum spin models.  We sample
$\P$ with the Metropolis algorithm.  The worm modifies a large cluster
of $O(L^3)$ $\D$ variables while conserving $\div \D$.  We define a
sweep as $3L^3$ Monte Carlo tries for $\P$ and one worm for $\D$.  We
measure $\beta S({\bf q})$ and determine $\epsilon({\bf q}) =
1/(1-\beta S({\bf q}))$. The results are plotted for several sets of
parameters in Figure~\ref{epsq}.  Passage of $\beta S(\bf q)$ through
$1$ gives rise to poles in the dielectric properties, leading to some
of the major qualitative features known in water: Firstly, a long
wavelength dielectric constant satisfying $\epsilon(0)>1$, secondly, a
band of wavevectors with negative dielectric constant, thirdly,
convergence of $\epsilon$ to $1$ for large $q$.

\begin{figure}[htb]
  \includegraphics[scale=.45] {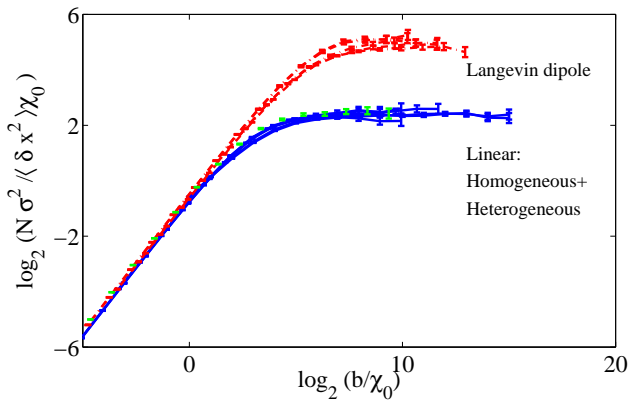}
  \caption{Scaled statistical error as a function of scaled blocking
    factor.  Curves for $L= 10, 13, 18$ and simulations of $N=2^{22}$
    sweeps.  Linear: eq.~(\ref{linear}), $\kappa=$ 1.0, 0.1, 0.03,
    0.015, $\kappa_t=0$, $\kappa_l=\kappa/8$.  Non-linear (blue):
    eq.~(\ref{nonlinear}).  $\gamma=$2, 1, 0.4, $\alpha=\kappa_l=0$
    scaled with $\chi_0$.  Heterogenous linear with $L=18$ where
    $\kappa=0.25$ for $0<z\le l/2$ and $\kappa=0.02$ for $L/2<z<L$
    scaled with $\chi_0=50$.  }
\label{blockfig}
\end{figure}

A central point of this letter is the demonstration of the
computational efficiency of our approach.  In the following we compare
a linear, non-local model eq.~(\ref{linear}), a heterogeneous, linear,
local model, and a non-linear model consisting of soft Langevin
dipoles eq.~(\ref{nonlinear}) We determine equilibration times with a
blocking method \cite{blocking}: Starting from $N=2^n$ recordings,
$x(t)$, one calculates an estimate of the mean and standard error for
blocks of data of length $b=2^m$ with $0\le m<n$.  We studied the
dynamics of variable $x(t)= \P^2(t,q=0)$, used to measure the $q=0$
dielectric constant through eq.~(\ref{k0}).  Blocking analysis leads
to a monotonically increasing estimate of the standard error, in
$\langle x \rangle$, $\sigma(b)$.  When $b$ the block size has reached
the autocorrelation time of the simulation the standard error
converges to $\sigma^2=2 \tau_{int} \langle \delta x^2 \rangle/N$,
where $\tau_{int}$ is the integrated autocorrelation time.  We
calculated the blocking curves for various values of system size and
model parameters, Figure~\ref{blockfig}.  Systems of different size
$L$ have blocking curves which superpose with no rescaling of the
data; the relaxation time (in sweeps) of the algorithm is independent
of the system size.  The cluster algorithm is indeed characterized by
a dynamic exponent $z=0$ and not hindered by the heterogeneous
material properties.
When we modify the dielectric properties we need to rescale both
horizontal and vertical axes to superpose data.  We find that the
scaling variables are $N \sigma^2(b) /\chi_0 \langle \delta x^2\rangle
$ as a function of $b/\chi_0$ where $\chi_0$ is the $q=0$
susceptibility; for the heterogeneous system we used the appropriate
$\chi_0$ for the large $\epsilon$ region.  Thus the simulation is
slowed by a factor $\chi_0$ when dielectric properties change. For
$\chi_0=1$ $\tau_{int}\sim 2$ sweeps when simulating with a simple
quadratic energy for the polarization fluctuations.  When using the
soft Langevin dipole the simulation is approximately six times slower.

The appearance of a long time scale in high dielectric media can be
understood as being due to the large ratio of the amplitude of
transverse to longitudinal fluctuations $2(1+K^{-1})= 2 \epsilon$,
which splits the experimental longitudinal and transverse relaxation
time scales \cite{kivelson}. This same splitting limits the efficiency
of our algorithm.  $O(\epsilon)$ sweeps are needed to fully
equilibrate the system and to generate the Keesom - van der Waals
interactions. Note, however, that the autocorrelation time of the
longitudinal modes which are important for interactions between
charges remains $O(1)$ sweeps even when the transverse and $q=0$ modes
are slow.

We conclude that our approach allows the investigation of electrostatic
interactions in heterogeneous, non-linear and non-local dielectric media. The
small computational costs for treating more complicated (elastic) polarization
energies are due to our local formulation of the problem.  Long ranged
interactions are not calculated explicitely, but generated through a
constrained energy functional for the electric field.  This is in marked
contrast to the conventional {\em global} solution of the Poisson
equation. The equivalent minimization of the functional
eq.~(\ref{dipoleenergy}) can be written as
\begin{math} 
  \P= (T+K)^{-1} \E_0
\end{math}. 
Since updating the {\em inverted} operator $(T+K)$ in 
computer simulations is very time consuming \cite{marchi},
most implicit solvent schemes rely on {\em approximate} solutions of the
{\em macroscopic} laws of electrostatics in heterogenous dielectric media
\cite{BashfordCase_arpc_00}.
Experience~\cite{igor,rottler} shows that the inherent advantages of the
formulation in terms of local fields are not lost in off-lattice
Molecular Dynamics implementations.  We therefore believe that the
present work opens the way to the development of more realistic implicit
solvents models for (bio)molecular simulations.

Collaboration financed by Volkswagenstiftung.

\bibliography{polar}

\begin{thebibliography}{23}
\expandafter\ifx\csname natexlab\endcsname\relax\def\natexlab#1{#1}\fi
\expandafter\ifx\csname bibnamefont\endcsname\relax
  \def\bibnamefont#1{#1}\fi
\expandafter\ifx\csname bibfnamefont\endcsname\relax
  \def\bibfnamefont#1{#1}\fi
\expandafter\ifx\csname citenamefont\endcsname\relax
  \def\citenamefont#1{#1}\fi
\expandafter\ifx\csname url\endcsname\relax
  \def\url#1{\texttt{#1}}\fi
\expandafter\ifx\csname urlprefix\endcsname\relax\def\urlprefix{URL }\fi
\providecommand{\bibinfo}[2]{#2}
\providecommand{\eprint}[2][]{\url{#2}}

\bibitem[{\citenamefont{Born}(1920)}]{born}
\bibinfo{author}{\bibfnamefont{M.}~\bibnamefont{Born}}, \bibinfo{journal}{Z.
  Phys} \textbf{\bibinfo{volume}{1}}, \bibinfo{pages}{45}
  (\bibinfo{year}{1920}).

\bibitem[{\citenamefont{Parsegian}(1969)}]{parsegian}
\bibinfo{author}{\bibfnamefont{A.}~\bibnamefont{Parsegian}},
  \bibinfo{journal}{Nature} \textbf{\bibinfo{volume}{221}},
  \bibinfo{pages}{844} (\bibinfo{year}{1969}).

\bibitem[{\citenamefont{Zhang et~al.}(2005)\citenamefont{Zhang, Kamenev, and
  Shklovskii}}]{sklovskii}
\bibinfo{author}{\bibfnamefont{J.}~\bibnamefont{Zhang}},
  \bibinfo{author}{\bibfnamefont{A.}~\bibnamefont{Kamenev}}, \bibnamefont{and}
  \bibinfo{author}{\bibfnamefont{B.}~\bibnamefont{Shklovskii}},
  \bibinfo{journal}{Phys. Rev. Lett.} \textbf{\bibinfo{volume}{95}},
  \bibinfo{pages}{148101} (\bibinfo{year}{2005}).

\bibitem[{\citenamefont{Dolgov et~al.}(1981)\citenamefont{Dolgov, Kirzhnits,
  and Maksimov}}]{dolgov}
\bibinfo{author}{\bibfnamefont{O.~V.} \bibnamefont{Dolgov}},
  \bibinfo{author}{\bibfnamefont{D.~A.} \bibnamefont{Kirzhnits}},
  \bibnamefont{and} \bibinfo{author}{\bibfnamefont{E.~G.}
  \bibnamefont{Maksimov}}, \bibinfo{journal}{Rev. Mod. Phys.}
  \textbf{\bibinfo{volume}{53}}, \bibinfo{pages}{81} (\bibinfo{year}{1981}).

\bibitem[{\citenamefont{Bopp et~al.}(1996)\citenamefont{Bopp, Kornyshev, and
  Sutmann}}]{bopp}
\bibinfo{author}{\bibfnamefont{P.~A.} \bibnamefont{Bopp}},
  \bibinfo{author}{\bibfnamefont{A.~A.} \bibnamefont{Kornyshev}},
  \bibnamefont{and} \bibinfo{author}{\bibfnamefont{G.}~\bibnamefont{Sutmann}},
  \bibinfo{journal}{Phys. Rev. Lett.} \textbf{\bibinfo{volume}{76}},
  \bibinfo{pages}{1280} (\bibinfo{year}{1996}).

\bibitem[{\citenamefont{Ball}(2003)}]{ball_nat_03}
\bibinfo{author}{\bibfnamefont{P.}~\bibnamefont{Ball}},
  \bibinfo{journal}{Nature} \textbf{\bibinfo{volume}{423}}, \bibinfo{pages}{25}
  (\bibinfo{year}{2003}).

\bibitem[{\citenamefont{Maggs}(2004)}]{acm1}
\bibinfo{author}{\bibfnamefont{A.~C.} \bibnamefont{Maggs}},
  \bibinfo{journal}{J. Chem. Phys.} \textbf{\bibinfo{volume}{120}},
  \bibinfo{pages}{3108} (\bibinfo{year}{2004}).

\bibitem[{\citenamefont{Marcus}(1956)}]{marcus}
\bibinfo{author}{\bibfnamefont{R.~A.} \bibnamefont{Marcus}},
  \bibinfo{journal}{J. Chem. Phys.} \textbf{\bibinfo{volume}{24}},
  \bibinfo{pages}{966} (\bibinfo{year}{1956}).

\bibitem[{\citenamefont{Kirzhnits}(1976)}]{kirzhnits}
\bibinfo{author}{\bibfnamefont{D.}~\bibnamefont{Kirzhnits}},
  \bibinfo{journal}{Uspekhi Fizicheskii Nauk} \textbf{\bibinfo{volume}{119}},
  \bibinfo{pages}{357} (\bibinfo{year}{1976}).

\bibitem[{\citenamefont{{L. Levrel \it et~al.}}(2005)}]{statphys}
\bibinfo{author}{\bibnamefont{{L. Levrel \it et~al.}}},
  \bibinfo{journal}{Pramana} \textbf{\bibinfo{volume}{64}},
  \bibinfo{pages}{1001} (\bibinfo{year}{2005}).

\bibitem[{\citenamefont{Swendsen and Wang}(1986)}]{swendson}
\bibinfo{author}{\bibfnamefont{R.~H.} \bibnamefont{Swendsen}} \bibnamefont{and}
  \bibinfo{author}{\bibfnamefont{J.-S.} \bibnamefont{Wang}},
  \bibinfo{journal}{Phys. Rev. Lett.} \textbf{\bibinfo{volume}{58}},
  \bibinfo{pages}{86} (\bibinfo{year}{1986}).

\bibitem[{\citenamefont{Alet and Sorensen}(2003)}]{alet}
\bibinfo{author}{\bibfnamefont{F.}~\bibnamefont{Alet}} \bibnamefont{and}
  \bibinfo{author}{\bibfnamefont{E.~S.} \bibnamefont{Sorensen}},
  \bibinfo{journal}{Phys. Rev. E} \textbf{\bibinfo{volume}{67}},
  \bibinfo{eid}{015701} (\bibinfo{year}{2003}).

\bibitem[{\citenamefont{{M. Marchi {\it et~al.}}}(2001)}]{marchi}
\bibinfo{author}{\bibnamefont{{M. Marchi {\it et~al.}}}}, \bibinfo{journal}{J.
  Chem. Phys.} \textbf{\bibinfo{volume}{114}}, \bibinfo{pages}{4377}
  (\bibinfo{year}{2001}).

\bibitem[{\citenamefont{Felderhof}(1977)}]{felderhof}
\bibinfo{author}{\bibfnamefont{B.~U.} \bibnamefont{Felderhof}},
  \bibinfo{journal}{J. Chem. Phys.} \textbf{\bibinfo{volume}{67}},
  \bibinfo{pages}{493} (\bibinfo{year}{1977}).

\bibitem[{\citenamefont{Fulton}(1975)}]{fulton}
\bibinfo{author}{\bibfnamefont{R.~L.} \bibnamefont{Fulton}},
  \bibinfo{journal}{J. Chem. Phys.} \textbf{\bibinfo{volume}{63}},
  \bibinfo{pages}{77} (\bibinfo{year}{1975}).

\bibitem[{\citenamefont{Kirkwood}(1939)}]{kirkwood}
\bibinfo{author}{\bibfnamefont{J.~G.} \bibnamefont{Kirkwood}},
  \bibinfo{journal}{J. Chem. Phys.} \textbf{\bibinfo{volume}{7}},
  \bibinfo{pages}{911} (\bibinfo{year}{1939}).

\bibitem[{\citenamefont{Kornyshev and Sutmann}(1997)}]{kornyshev}
\bibinfo{author}{\bibfnamefont{A.~A.} \bibnamefont{Kornyshev}}
  \bibnamefont{and} \bibinfo{author}{\bibfnamefont{G.}~\bibnamefont{Sutmann}},
  \bibinfo{journal}{Phys. Rev. Lett.} \textbf{\bibinfo{volume}{79}},
  \bibinfo{pages}{3435} (\bibinfo{year}{1997}).

\bibitem[{\citenamefont{Florian and Warshel}(1997)}]{warshel2}
\bibinfo{author}{\bibfnamefont{J.}~\bibnamefont{Florian}} \bibnamefont{and}
  \bibinfo{author}{\bibfnamefont{A.}~\bibnamefont{Warshel}},
  \bibinfo{journal}{J. Phys. Chem. B} \textbf{\bibinfo{volume}{101}},
  \bibinfo{pages}{5583} (\bibinfo{year}{1997}).

\bibitem[{\citenamefont{Flyvbjerg and Petersen}(1989)}]{blocking}
\bibinfo{author}{\bibfnamefont{H.}~\bibnamefont{Flyvbjerg}} \bibnamefont{and}
  \bibinfo{author}{\bibfnamefont{H.~G.} \bibnamefont{Petersen}},
  \bibinfo{journal}{J. Chem. Phys.} \textbf{\bibinfo{volume}{91}},
  \bibinfo{pages}{461} (\bibinfo{year}{1989}).

\bibitem[{\citenamefont{Kivelson and Friedman}(1989)}]{kivelson}
\bibinfo{author}{\bibfnamefont{D.}~\bibnamefont{Kivelson}} \bibnamefont{and}
  \bibinfo{author}{\bibfnamefont{H.}~\bibnamefont{Friedman}},
  \bibinfo{journal}{J. Phys. Chem.} \textbf{\bibinfo{volume}{93}},
  \bibinfo{pages}{7026} (\bibinfo{year}{1989}).

\bibitem[{\citenamefont{Bashford and Case}(2000)}]{BashfordCase_arpc_00}
\bibinfo{author}{\bibfnamefont{D.}~\bibnamefont{Bashford}} \bibnamefont{and}
  \bibinfo{author}{\bibfnamefont{D.}~\bibnamefont{Case}}, \bibinfo{journal}{An.
  Rev. Phys. Chem.} \textbf{\bibinfo{volume}{51}}, \bibinfo{pages}{129}
  (\bibinfo{year}{2000}).

\bibitem[{\citenamefont{Pasichnyk and D\"{u}nweg}(2004)}]{igor}
\bibinfo{author}{\bibfnamefont{I.}~\bibnamefont{Pasichnyk}} \bibnamefont{and}
  \bibinfo{author}{\bibfnamefont{B.}~\bibnamefont{D\"{u}nweg}},
  \bibinfo{journal}{J. Phys. Cond. Mat.} \textbf{\bibinfo{volume}{16}},
  \bibinfo{pages}{S3999} (\bibinfo{year}{2004}).

\bibitem[{\citenamefont{Rottler and Maggs}(2004)}]{rottler}
\bibinfo{author}{\bibfnamefont{J.}~\bibnamefont{Rottler}} \bibnamefont{and}
  \bibinfo{author}{\bibfnamefont{A.~C.} \bibnamefont{Maggs}},
  \bibinfo{journal}{Phys. Rev. Lett.} \textbf{\bibinfo{volume}{93}},
  \bibinfo{eid}{170201} (\bibinfo{year}{2004}).

\end{thebibliography}
\end{document}